\documentclass[prx, twocolumn,longbibliography]{revtex4-1}
\usepackage{float}
\usepackage{bbm}
\usepackage{epsfig}
\usepackage{epstopdf}
\usepackage{graphicx, subfigure}
\usepackage{amsmath,amssymb}
\usepackage{color}
\usepackage{hyperref}
\usepackage[capitalize]{cleveref}
\usepackage{braket,mleftright}

\begin{document}

\title{Fast high-fidelity entangling gates for spin qubits in Si double quantum dots }
\author{F.~A.~Calderon-Vargas$^1$}
\email{f.calderon@vt.edu}
\author{George~S.~Barron$^1$}
\author{Xiu-Hao Deng$^1$}
\author{A.~J.~Sigillito$^2$}
\author{Edwin~Barnes$^1$}
\email{efbarnes@vt.edu}
\author{Sophia~E.~Economou$^1$}
\email{economou@vt.edu}
\affiliation{$^1$Department of Physics, Virginia Tech, Blacksburg, Virginia 24061, USA\\ $^2$Department of Physics, Princeton University, Princeton, New Jersey 08544, USA}

\begin{abstract}
Implementing high-fidelity two-qubit gates in single-electron spin qubits in silicon double quantum dots is still a major challenge. In this work, we employ analytical methods to design control pulses that generate high-fidelity entangling gates for quantum computers based on this platform. Using realistic parameters and initially assuming a noise-free environment, we present simple control pulses that generate {\sc cnot}, {\sc cphase}, and {\sc cz} gates with average fidelities greater than $99.99\%$ and gate times as short as $45\mathrm{ns}$. Moreover, using the local invariants of the system's evolution operator, we show that a simple square pulse generates a {\sc cnot} gate in less than 27 ns and with a fidelity greater than 99.99\%. Last, we use the same analytical methods to generate two-qubit gates locally equivalent to $\sqrt{\textsc{cnot}}$ and $\sqrt{\textsc{cz}}$ that are used to implement simple two-piece pulse sequences that produce high-fidelity {\sc cnot} and {\sc cz} gates in the presence of low-frequency noise.  
\end{abstract}


\maketitle

\section{Introduction}\label{sec:Intro}
Among the many candidates to become the hardware of the first quantum computer, single electron spin qubits in silicon quantum dots (QDs) present a distinctive potential for scalability due to the well-established fabrication technologies and all-electrical control~\cite{Loss1998,Zwanenburg2013}. In the last five years, several groups have demonstrated high-fidelity single-qubit gates~\cite{Veldhorst2014,Kawakami2016,Takeda2016,Yoneda2017,Zajac2018} with fidelities higher than 99.9\%~\cite{Yoneda2017}. Two-qubit gates---a second crucial component of any quantum logic circuit---have also been demonstrated in these systems~\cite{Brunner2011,Veldhorst2015a,Watson2018,Zajac2018,Huang2018c,Xue2018a}, however the fidelities in this case have not yet exceeded 98\%~\cite{Huang2018c}. This is mostly due to charge noise that causes fluctuations in the tunneling and detuning energies~\cite{Huang2018a,Dijk2018}, which directly affects the exchange coupling and single-qubit resonance frequencies~\cite{Chan2018,Yoneda2017}. A second source of decoherence comes from fluctuations in the nuclear spin bath of the surrounding substrate and its hyperfine interaction with the electron spins, but for natural silicon, in contrast to other semiconductor platforms, the abundance of the nuclear-spin isotope $^{29}\mathrm{Si}$ is only $4.7\%$. Moreover, the spinful isotopes can be removed through isotopic purification~\cite{Itoh2014}, largely eliminating this type of noise.

Therefore, charge noise remains the main obstacle in the path towards high-fidelity gates. Several methods have been proposed to mitigate its effect on qubit operations. In particular, it has been experimentally shown that the exchange energy's sensitivity to charge noise can be reduced to second order by symmetric operations~\cite{Reed2016,Martins2016,Zhang2017,Zajac2018}. However, the remaining noise still limits gate fidelities, necessitating the use of additional techniques. One approach is dynamically corrected gates (DCGs), where pulse shaping or carefully designed pulse sequences are used to suppress the effects of noise during operations. Most DCG schemes work only if the noise is concentrated at low frequencies. This is true of charge noise in Si quantum dots, where the power spectrum has been measured to have the form $1/f^\alpha$, with $\alpha\approx1$~\cite{Yoneda2017,Chan2018}. In fact, DCGs have been proposed for both single- and two-qubit gates~\cite{Zeng2018,Russ2017,Gungordu2018}. However, it is important to make the pulses as fast as possible in these schemes, because the assumption of slow noise becomes invalid for longer gate times, weakening the power of DCGs. Of course, speeding up gates is generally desirable as it leads to faster quantum information processing and improves fidelities in the presence of finite coherence times.

The speed of two-qubit gates in exchange-coupled quantum dots is ultimately set by the strength of the exchange interaction. However, most proposed two-qubit gates take longer than what one would expect just based on the interaction strength. Recently, it has been experimentally demonstrated that entangling gates can be generated in a single shot using electric dipole spin resonance (EDSR) in conjunction with an exchange pulse~\cite{Zajac2018,Russ2017}. However, this approach requires the pulse to be slow compared to the exchange time scale in order to avoid unwanted transitions in the spectrum and to accumulate the correct phases for the desired entangling gate, reducing the effectiveness of dynamical noise suppression and leading to fidelity loss.

In this work, we show that fast, high-fidelity two-qubit entangling gates in Si quantum dots can be achieved with analytical pulse-shaping techniques. Our gate designs are based on single-shot exchange pulses combined with EDSR control. We overcome naive gate speed limits by purposely driving spectrally close unwanted transitions within the logical space and incorporating them into our gate design rather than simply avoiding them~\cite{Economou2006,Economou2012,Economou2015}. In this way, we achieve entangling operations with fidelities exceeding 99.99\% in the absence of noise in times as low as 45 ns, a factor of 2 improvement over previous methods. Moreover, we calculate the local invariants~\cite{Makhlin2002} of the system's evolution operator and find that a simple square pulse generates, up to local operations, a {\sc cnot} gate in less than 27 ns and with a fidelity larger than $99.99\%$ in the absence of noise. Finally, we combine our pulse-shaping methods with dynamical gate correction to further improve fidelities. We design fast, robust pulses for the three most commonly used entangling operations: {\sc cnot}, {\sc cz} and {\sc cphase} gates, which arise in numerous algorithms pertaining to quantum computing and simulation.

The paper is organized as follows. In Sec.~\ref{sec:two-qubit Hamiltonian}, we introduce the two-qubit Hamiltonian. Then, in Sec.~\ref{sec:cnot gate}, we present a simple analytical method, based on a general analytical solution for the unitary evolution of an arbitrary two-level system, to design control pulses that generate high-fidelity {\sc cnot} gates. In Sec.~\ref{sec:square pulse}, we compute the local invariants of the evolution operator and show that a short square pulse generates a high-fidelity {\sc cnot} gate. In Sec.~\ref{sec:cphase and cz gates}, we show that with a simple hyperbolic secant (sech) function as the control pulse we can generate high-fidelity {\sc $\theta$-cphase} gates, with arbitrary phase angle, including the {\sc cz} (or {\sc $\pi$-cphase}) gate. We conclude in Sec.~\ref{sec:Conclusion}.

\section{Two-qubit Hamiltonian}\label{sec:two-qubit Hamiltonian}
\begin{figure}[t]
  \centering
  \includegraphics[trim=4cm 4cm 0cm 3cm, clip=true,width=10.5cm, angle=0]{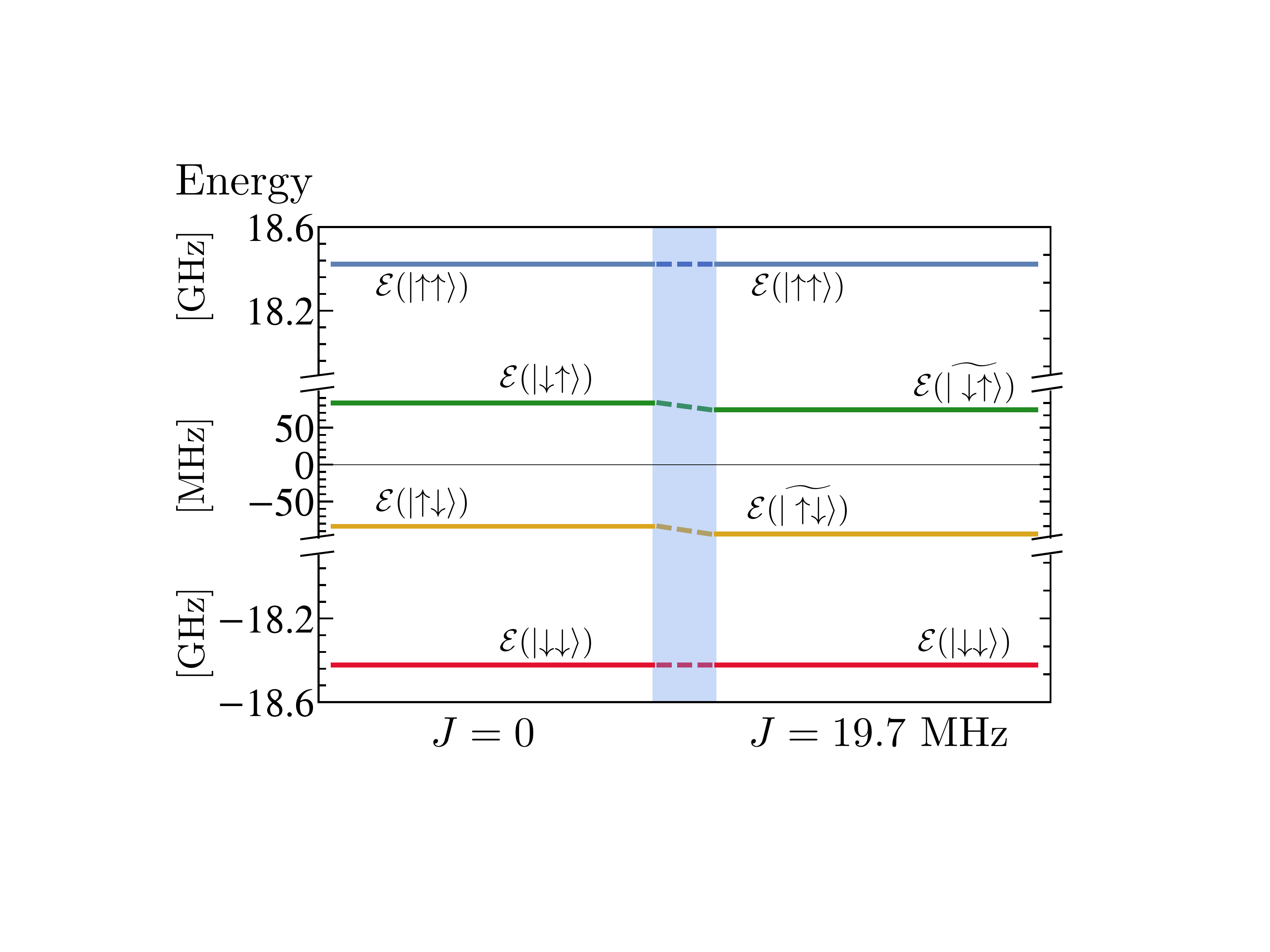}
  \caption{Eigenenergies of the system's Hamiltonian, in the absence of AC driving, for zero exchange coupling ($J=0$, left side), and for nonzero exchange coupling ($J=19.7\mathrm{MHz}$, right side). We use the parameters reported in Ref.~\onlinecite{Zajac2018} to calculate the magnitude of the four eigenenergies. Note that when the exchange coupling is turned on the energies of the states with antiparallel spin are lowered by $\sim J/2$ for $J\ll(B_{z,R}-B_{z,L})$, and thus the conditional spin flips (i.e. the ESR frequency of one spin depends on the other spin state) are now distiguishable from each other.} \label{fig:1}
\end{figure}

The system being studied comprises two electrons confined in a silicon double quantum dot (DQD). The two-electron system is under an external magnetic field, applied on the $z$-axis, and an inhomogenous magnetic field, created by a nearby micromagnet, with large gradients along the $z$- and $x$-axes (the two-dimensional electron gas is in the $z$-$x$ plane and the two quantum dots are spatially separated along the $x$-direction). In particular, the component of the magnetic field out of the plane, $B_y$, varies along the $z$ direction, while $B_z$ varies along the $x$ direction. The spatial dependence of $B_y$ enables EDSR control in which the application of a microwave pulse to a nearby metal gate rapidly oscillates the position of the electrons in the $z$-direction, causing a transverse oscillating magnetic field $B_y$ in the electron's frame of reference. The gradient of $B_z$ along the $x$ direction gives the two electrons distinct resonance frequencies, allowing individual addressability. Following the experimental setup reported in Ref.~\onlinecite{Zajac2018}, we assume that the system is operated at a symmetric operation point where the exchange coupling $J(t)$ is first-order insensitive to electric noise~\cite{Martins2016,Reed2016}. The exchange coupling is controlled by a middle metal gate that modulates the energy barrier between dots and, therefore, the interdot tunneling amplitude. 

We use the Heisenberg Hamiltonian to model the system,
\begin{equation}\label{eq:Heisenberg Hamiltonian}
H(t)=J(t)(\mathbf{S}_L\cdot\mathbf{S}_R-1/4)+\mathbf{S}_L\cdot\mathbf{B}_L+\mathbf{S}_R\cdot\mathbf{B}_R,
\end{equation}
where $\mathbf{S}_L$($\mathbf{S}_R$) is the spin of the electron in the left(right) quantum dot, $ \mathbf{B}_L=(0,B_{y,L}(t),B_z^{ext}+B_{z,L}(t))^T$ and $ \mathbf{B}_R=(0,B_{y,R}(t),B_z^{ext}+B_{z,R}(t))^T$ are the magnetic fields (in energy units) on the spins in the left and right dots, respectively. Here, $B_z^{ext}$ is the homogeneous external magnetic field in the $z$-direction, $B_{z,L}(t)$($B_{z,R}(t)$) is the local magnetic field in the left(right) dot induced by the micromagnet's inhomogeneous field, and $B_{y,q}(t)=B_{y,q}^0+B_{y,q}^1 \cos(\omega t+\phi)$, ($q=L,R$), is the transverse oscillating field composed of a static part $B_{y,q}^0$ and a variable amplitude $B_{y,q}^1$ controlled by a nearby metal gate. In similar fashion, the $z$-component of the local magnetic field has a time-dependent component, i.e. $B_{z,q}(t)=B_{z,q}^0+B_{z,q}^1(t)$, caused by the change in the position of the electrons in the magnetic field gradient when the energy barrier between dots is changed by the middle metal gate. Therefore, when the exchange coupling is turned on, the two parameters $B_{z,L}^1$ and $B_{z,R}^1$ cause shifts in the single-qubit resonance frequencies. Following Ref.~\onlinecite{Zajac2018}, our analysis is focused on the regime where the Zeeman splitting  of the two qubits is much larger than the strength of the exchange interaction $J\ll (B_{z,R}-B_{z,L})$. In particular, in the regime where the exchange interaction is negligible, $J\approx 0$ (see Fig.~\ref{fig:1}), single-qubit gates can be performed by matching the drive frequency $\omega$ with the resonance frequency of either dot, and by adjusting the drive phase $\phi$ one can control the single-qubit rotation axis in the Bloch sphere's equatorial plane.

In the absence of AC driving, the Hamiltonian \eqref{eq:Heisenberg Hamiltonian} in the diabatic two-qubit basis $\{\ket{\uparrow\uparrow}, \ket{\downarrow\uparrow}, \ket{\uparrow\downarrow}, \ket{\downarrow\downarrow}\}$ is:
\begin{widetext}
\begin{equation}\label{eq:H_0}
H_0(t)=\left(
\begin{array}{cccc}
 E_z+E_z^1[J] & 0 & 0 & 0 \\
 0 & \frac{1}{2}(\Delta E_z+\Delta E_z^1[J]-J(t)) & \frac{J(t)}{2} & 0 \\
0 & \frac{J(t)}{2} & \frac{1}{2} (-\Delta E_z-\Delta E_z^1[J]-J(t)) & 0 \\
 0 & 0 & 0 & -E_z -E_z^1[J]\\
\end{array}
\right),
\end{equation}
\end{widetext}
where $E_z=B_z^{ext}+(B_{z,R}^0+B_{z,L}^0)/2$ is the average Zeeman splitting, $\Delta E_z=B_{z,R}^0-B_{z,L}^0$ is the Zeeman splitting between dots, and $E_z^1[J]=(B_{z,R}^1+B_{z,L}^1)/2$ and $\Delta E_z^1[J]= B_{z,R}^1-B_{z,L}^1$ represent the Zeeman shifts that occur when the exchange $J$ is turned on. Here $E_z^1[J]$ and $\Delta E_z^1[J]$ would vanish if $J$ vanishes too. If the exchange is adiabatically turned on (relative to the Zeeman splitting of the two qubits), in the absence of AC driving, we can use the Hamiltonian's instantaneous adiabatic eigenstates as the new basis. Accordingly, the Hamiltonian's instantaneous eigenvalues are
\begin{equation}\label{eq:E(up,up)}
\mathcal{E}(\ket{\uparrow\uparrow})=E_z+E_z^1, 
\end{equation}
\begin{equation} \label{eq:E(down,up)}
\mathcal{E}(\widetilde{\ket{\downarrow\uparrow}})=\frac{1}{2}(-J +\sqrt{(\Delta E_z+\Delta E_z^1)^2+J^2}), 
\end{equation}
\begin{equation}\label{eq:E(up,down)}
\mathcal{E}(\widetilde{\ket{\uparrow\downarrow}})=\frac{1}{2}(-J -\sqrt{(\Delta E_z+\Delta E_z^1)^2+J^2}), 
\end{equation}
\begin{equation}\label{eq:E(down,down)}
\mathcal{E}(\ket{\downarrow\downarrow})=-E_z-E_z^1,
\end{equation}
where $\widetilde{\ket{\downarrow\uparrow}},\widetilde{\ket{\uparrow\downarrow}} $ are hybridizations of the $ \ket{\downarrow\uparrow}, \ket{\uparrow\downarrow}$ spin states.

The fast high-fidelity two-qubit gates proposed in this work require microwave AC driving along with a nonzero constant exchange interaction. Since we are assuming that $J\ll \Delta E_z+\Delta E_z^1$, the Hamiltonian \eqref{eq:Heisenberg Hamiltonian} in the instantaneous adiabatic basis $\{\ket{\uparrow\uparrow}, \widetilde{\ket{\downarrow\uparrow}}, \widetilde{\ket{\uparrow\downarrow}}, \ket{\downarrow\downarrow} \}$ can be simplified by expanding $\sqrt{(\Delta E_z+\Delta E_z^1)^2+J^2}\approx\Delta E_z+\Delta E_z^1+ J^2/2(\Delta E_z+\Delta E_z^1)$. Moreover, given that the Zeeman splitting is much larger than the exchange coupling, we only keep terms up to leading order in $J/(\Delta E_z+\Delta E_z^1)$. At this point it is convenient to switch to the interaction picture given by $H_0(J=0)$, Eq.~\eqref{eq:H_0}, because we do not include the free, uncoupled evolution in the definition of our gates---they are generated purely by AC driving. Our results can easily be adapted if one wishes to define logical gates with respect to a different frame. Consequently, in the adiabatic basis, the interaction picture Hamiltonian is:
\begin{widetext}
\begin{equation}\label{eq:Hamiltonian_int}
H_{int}=\left(
\begin{array}{cccc}
 E_z^1 & -i \frac{B_{y,L}\Delta_+^{(1)}}{2}e^{-\frac{i}{2}  (\Delta E_z - 2 E_z) t} &  -i \frac{B_{y,R}\Delta_-^{(2)}}{2}e^{\frac{i}{2}  (\Delta E_z + 2 E_z) t} & 0 \\
  i \frac{B_{y,L}\Delta_+^{(1)}}{2}e^{\frac{i}{2}  (\Delta E_z - 2 E_z) t} & \frac{1}{2}\left(\Delta E_z^1-J+\frac{J^2}{2(\Delta E_z+\Delta E_z^1)} \right) & 0 & -i\frac{B_{y,R}\Delta_+^{(2)}}{2} e^{\frac{i}{2}  (\Delta E_z + 2 E_z) t} \\
  i \frac{B_{y,R}\Delta_-^{(2)}}{2}e^{-\frac{i}{2}  (\Delta E_z + 2 E_z) t}& 0 & -\frac{1}{2}\left(\Delta E_z^1+J+\frac{J^2}{2(\Delta E_z+\Delta E_z^1)} \right) &  -i\frac{B_{y,L}\Delta_-^{(1)}}{2} e^{-\frac{i}{2}  (\Delta E_z - 2 E_z) t} \\
 0 & i\frac{B_{y,R}\Delta_+^{(2)}}{2} e^{-\frac{i}{2}  (\Delta E_z + 2 E_z) t} &  i\frac{B_{y,L}\Delta_-^{(1)}}{2} e^{\frac{i}{2}  (\Delta E_z - 2 E_z) t} &  -E_z^1 \\
\end{array}
\right),
\end{equation}
\end{widetext}
where $\Delta_\pm^{(1)}(t)=1\pm J B_{y,R}(t)/2B_{y,L}(t)(\Delta E_z+\Delta E_z^1)$ and $\Delta_\pm^{(2)}(t)=1\pm J B_{y,L}(t)/2B_{y,R}(t)(\Delta E_z+\Delta E_z^1)$. Our goal is to find pulse envelopes $B_{y,L}^1(t)$, $B_{y,R}^1(t)$ that render the corresponding evolution operator $U_{int}$ locally equivalent to a two-qubit entangling gate at the end of the pulse. Note that although we have expressed $H_{int}$ in the adiabatic basis, the gates we design apply equally well to the original diabatic basis since the two bases coincide at the end of the gate when the exchange coupling is switched off.

A Hamiltonian similar to the one in Eq.~\ref{eq:Hamiltonian_int} has been addressed in a previous work on superconducting qubits~\cite{Economou2015}, where the SWIPHT (speeding up wave forms by inducing phases to harmful transitions) protocol was introduced and control pulses were designed for the {\sc cnot} and {\sc cz} gates. The analytical methods developed in the aforementioned work and earlier works~\cite{Barnes2013,Economou2012,Barnes2012a,Economou2006} are the basis for the control pulses we present in the following sections.

For the numerical results hereafter, we use the parameters reported in Ref.~\onlinecite{Zajac2018}: $B_z^{ext}/2\pi=14\mathrm{GHz}$, $\Delta E_z/2\pi=214\mathrm{MHz}$, $B_{z,L}^0/2\pi=4.287\mathrm{GHz}$, $B_{y,L}^0/2\pi=5\mathrm{MHz}$, $B_{y,R}^0/2\pi=55\mathrm{MHz}$, $J/2\pi=19.7\mathrm{MHz}$, $B_{y,R}^1(t)=B_{y,L}^1(t)$, $\phi=3\pi/2 $, $E_z^1/2\pi=29.23\mathrm{MHz} $, $\Delta E_z^1/2\pi= -46.94\mathrm{MHz}$.

\section{{\sc cnot} gate from partial reverse-engineering}\label{sec:cnot gate}
\begin{figure}[!tbp]
  \centering
  \includegraphics[trim=3.2cm 0cm 3cm 0cm, clip=true,width=9cm, angle=0]{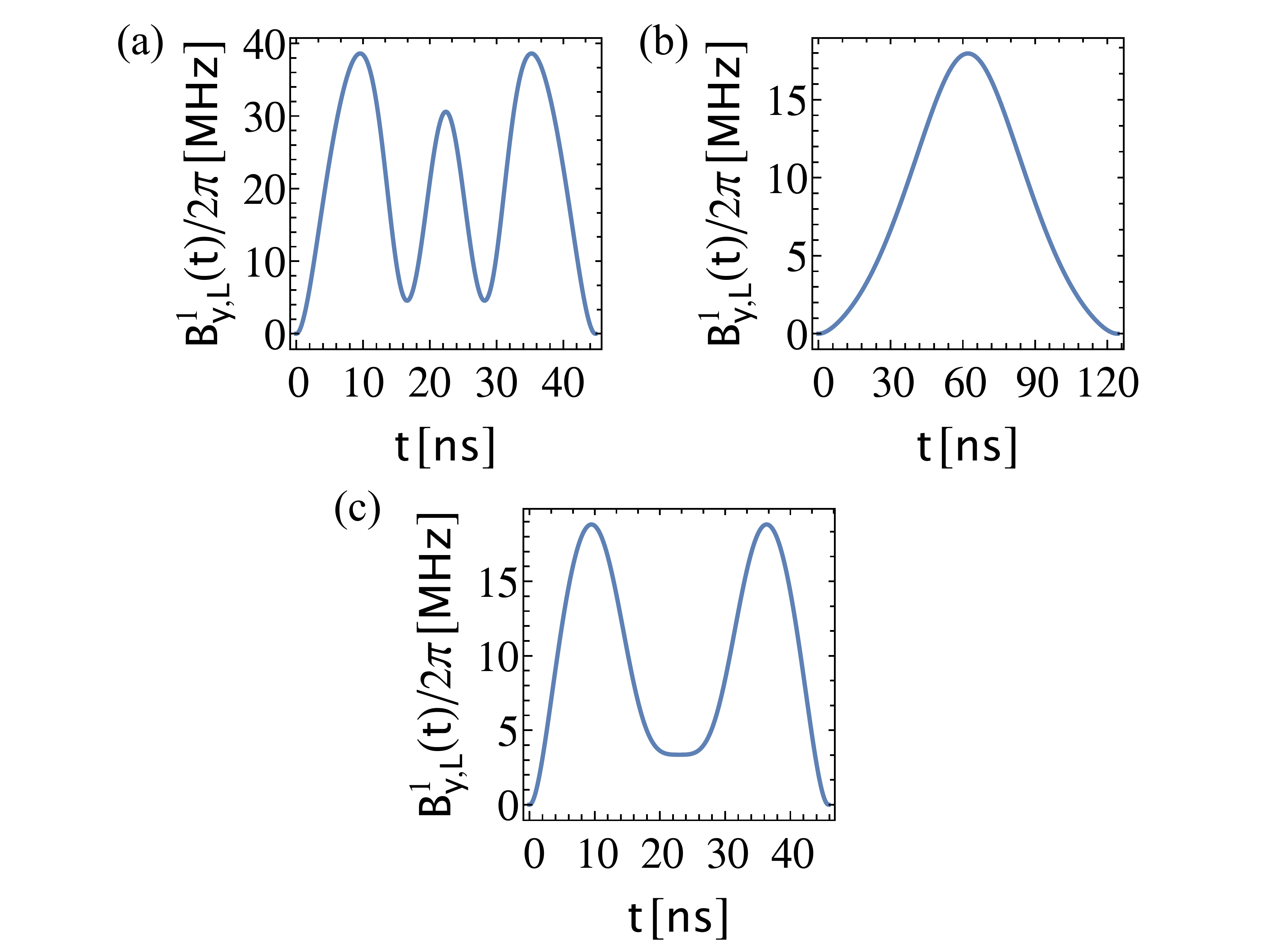}
  \caption{(a,b) Control pulse that generates a maximally entangling gate locally equivalent to {\sc cnot} in (a) $~45\mathrm{ns}$ with $99.994\%$ fidelity, and (b) $~124\mathrm{ns}$ with $99.997\%$ fidelity. (c) Control pulse that generates a two-qubit gate locally equivalent to $\sqrt{\text{{\sc cnot}}}$ in $~46\mathrm{ns}$. The two-piece pulse is shown in Eq.~\eqref{eq:two-piece-cnot} and generates a {\sc cnot} gate with 99.9999\% fidelity.}\label{fig:2}
\end{figure}

In order to facilitate the analytical derivation of our shaped pulses, it is convenient to momentarily move the Hamiltonian \eqref{eq:Heisenberg Hamiltonian} (in the adiabatic basis) to a rotating frame and apply the rotating wave approximation. Since the interaction and rotating frames are related by local unitaries, if we design a pulse that implements an entangling gate in the rotating frame, it will implement the same gate in the interaction frame up to these local unitaries. We stress that although we make the rotating wave approximation when designing pulses, we do not make this approximation when we numerically calculate the fidelity of the corresponding gates. Accordingly, moving the Hamiltonian \eqref{eq:Heisenberg Hamiltonian} to the rotating frame, ${H_{rot}(t)=U_{\omega}H(t)U_{\omega}^{\dagger}-iU_{\omega}\dot{U_{\omega}}^{\dagger}}$ with $U_{\omega}=\exp[i \omega t (S_{z,R}+S_{z,L})/\hbar]$, and applying the rotating wave approximation, we get in the adiabatic basis for $\phi=3\pi/2$:
\begin{widetext}
\begin{equation}\label{eq:Hamiltonian_rot}
H_{rot}=\left(
\begin{array}{cccc}
 E_z+E_z^1-\omega & \frac{B_{y,L}^1\delta_+^{(1)}}{4} & \frac{B_{y,R}^1\delta_-^{(2)}}{4} & 0 \\
 \frac{B_{y,L}^1\delta_+^{(1)}}{4} & \frac{1}{2}\left(\Delta E_z+\Delta E_z^1-J+\frac{J^2}{2(\Delta E_z+\Delta E_z^1)} \right) & 0 & \frac{B_{y,R}^1\delta_+^{(2)}}{4} \\
 \frac{B_{y,R}^1\delta_-^{(2)}}{4} & 0 & -\frac{1}{2}\left(\Delta E_z+\Delta E_z^1+J+\frac{J^2}{2(\Delta E_z+\Delta E_z^1)} \right) & \frac{B_{y,L}^1\delta_-^{(1)}}{4} \\
 0 & \frac{B_{y,R}^1\delta_+^{(2)}}{4} & \frac{B_{y,L}^1\delta_-^{(1)}}{4} & \omega -E_z-E_z^1 \\ 
\end{array}
\right),
\end{equation}
\end{widetext}
where $\delta_\pm^{(1)}=1\pm J B_{y,R}^1/2B_{y,L}^1(\Delta E_z+\Delta E_z^1)$ and $\delta_\pm^{(2)}=1\pm J B_{y,L}^1/2B_{y,R}^1(\Delta E_z+\Delta E_z^1)$. If the driving frequency $\omega$ is near to the left qubit's resonance frequency, $\vert\mathcal{E}(\ket{\uparrow\uparrow})-\mathcal{E}(\widetilde{\ket{\downarrow\uparrow}})\vert$, the terms $B_{y,R}^1\delta_\pm^{(2)}/4$ in $H_{rot}$ \eqref{eq:Hamiltonian_rot} can be dropped because $B_{y,R}^1\delta_\pm^{(2)}/4\ll \Delta E_z+\Delta E_z^1$. As a consequence, we have two  blocks along the diagonal, $S_1$ and $S_2$, spanned by $\{\ket{\uparrow\uparrow}, \widetilde{\ket{\downarrow\uparrow}} \}$ and $\{ \widetilde{\ket{\uparrow\downarrow}}, \ket{\downarrow\downarrow}\}$, respectively, which are effectively decoupled in the rotating Hamiltonian \eqref{eq:Hamiltonian_rot} and are separated in energy by $J$. The resonance frequency of each block is $\omega_1=E_z+E_z^1-(\Delta E_z+\Delta E_z^1-J+J^2/2(\Delta E_z+\Delta E_z^1))/2$ and  $\omega_2=E_z+E_z^1-(\Delta E_z+\Delta E_z^1+J+J^2/2(\Delta E_z+\Delta E_z^1))/2$, respectively.  We start by matching the driving frequency $\omega$ with the resonance frequency of $S_1$, i.e. inducing a resonant spin flip between the states $\ket{\uparrow\uparrow}$ and $\widetilde{\ket{\downarrow\uparrow}}$. Moreover, since $J\ll\Delta E_z+\Delta E_z^1$ and $B_{y,R}^1(t)=B_{y,L}^1(t)$ we set $\delta_\pm^{(1)}\approx1$. Consequently, the Hamiltonian \eqref{eq:Hamiltonian_rot} becomes, up to a local phase in $S_2$,
\begin{equation}\label{eq:Hamiltonian_rot_with_only_diag_block}
H_{rot}=\left(
\begin{array}{cccc}
 0 & \frac{B_{y,L}^1(t)}{4} & 0 & 0 \\
 \frac{B_{y,L}^1(t)}{4} & 0 & 0 & 0 \\
 0 & 0 & -\frac{J}{2} & \frac{B_{y,L}^1(t)}{4} \\
 0 & 0 & \frac{B_{y,L}^1(t)}{4} & \frac{J}{2} \\
\end{array}
\right).
\end{equation}
The block $S_1$ can be readily integrated to give the evolution operator $U_{S_1}(t)=\exp\{-\tfrac{i}{4}\int_0^t dt' B_{y,L}(t')\sigma_x\}$. For the block $S_2$, in Refs.~\onlinecite{Economou2015,Barnes2013} it has been shown that for a two-level system with detuning $\Delta$ and driving field $\Omega(t)$ the system's evolution operator $\mathcal{U}(t)$ and pulse $\Omega(t)$ can be expressed in terms of a single real function $\chi(t)$:
\begin{align}
\mathcal{U}(t)&=e^{-i\frac{\pi}{4}\sigma_y}\left(
\begin{array}{cc}
 \cos\chi e^{i\psi_-} & \sin\chi e^{-i\psi_+} \\
-\sin\chi e^{i\psi_+} & \cos\chi e^{-i\psi_-}  \\
\end{array}
\right),\label{eq:two-level parametrized operator}\\
\Omega(t)&=\frac{\ddot{\chi}}{2\sqrt{\frac{\Delta^2}{4}-\dot{\chi}^2}}-\sqrt{\frac{\Delta^2}{4}-\dot{\chi}^2}\cot (2\chi),\label{eq:Omega(t)}
\end{align}
where $\psi_{\pm}(t)=\int^t_0 dt'\sqrt{\Delta^2/4 -\dot{\chi}^2(t')}\csc[2\chi(t')]\pm\arcsin[2\dot{\chi}(t)/\Delta]/2$, and any choice of $\chi(t)$ that fulfills $\vert\dot{\chi}\vert\leqslant\vert\Delta/2\vert$ gives an exact solution to the system's Schr\"odinger equation. We have included a $\pi/2$-rotation about $y$ in Eq.~\eqref{eq:two-level parametrized operator} to account for the fact that, unlike in the present case, the driving field appears in the diagonal entries of the Hamiltonian in Ref.~\cite{Barnes2013}. We can apply this method to the block $S_2$ in Eq.~\ref{eq:Hamiltonian_rot_with_only_diag_block}, where $B_{y,L}^1(t)=4\Omega(t)$ and $J=\Delta$. Following Ref.~\onlinecite{Economou2015}, in order to generate a $\pi$ $x$-rotation in $S_1$ and an identity operation in $S_2$, which would correspond to a {\sc cnot} gate, the following conditions must be satisfied: $\int^{\tau}_0 B_{y,L}^1(t)=2\pi$, $\chi(0)=\chi(\tau)=\pi/4$, and $\dot{\chi}(0)=\dot{\chi}(\tau)=0$, where $\tau$ is the pulse duration (gate time).  For this particular gate, the phases $\psi_{\pm}$ would only affect the local unitaries that accompany the entangling operation, and thus no constraint is imposed. An ansatz for $\chi(t)$ that automatically satisfies the previous conditions is
\begin{equation}\label{eq:ansatz}
\chi(t)=A(\frac{t}{\tau})^4(1-\frac{t}{\tau})^4+\frac{\pi}{4}.
\end{equation}
The parameters $A$ and $\tau$ in Eq.~\eqref{eq:ansatz} can be adjusted to generate {\sc cnot} gates with different gate times and pulse amplitudes. Figures~\hyperref[fig:2]{\ref*{fig:2}(a)} and~\hyperref[fig:2]{\ref*{fig:2}(b)} show a pair of control pulses, $B_{y,L}^1(t)=4\Omega(t)$ with $\Omega(t)$ given by Eq.~\ref{eq:Omega(t)}, that generate {\sc cnot} gates. The parameters that produce the control pulse in Fig.~\hyperref[fig:2]{\ref*{fig:2}(a)} are $A_{(a)}=139.2947$ and $\tau_{(a)}  =5.54498/J$, and the parameters that produce the second pulse, Fig.~\hyperref[fig:2]{\ref*{fig:2}(b)}, are $A_{(b)}=61.4617$ and $\tau_{(b)}=15.38016/J$. Therefore we get a {\sc cnot} gate in the interaction picture:
\begin{equation}\label{eq:cnot}
\left(\begin{smallmatrix}
  0 & 1 & 0 & 0 \\
  1 & 0 & 0 & 0 \\
  0 & 0 & 1 & 0 \\
  0 & 0 & 0 & 1
\end{smallmatrix}\right)\approx K_1^{(j)}U_{int}(\tau_{(j)}) K_2^{(j)},
\end{equation}
where $j=a,b$ identifies the set of $\chi$-parameters used \{$A_{(j)},\tau_{(j)}$\}, $\omega=\omega_1$, and $K_i^{(j)}$ are Kronecker products of single-qubit gates (see Appendix \ref{app:local gates k1, k2}), where the superscript in $K_i^{(j)}$ refers to which pulse we are using to generate the {\sc cnot} (see Fig.~\ref{fig:2}). This distinction is needed since different pulses will produce different sets of single-qubit gates. Using the gate fidelity given by~\cite{Pedersen2007}:
\begin{equation}\label{eq:Fidelity_equation}
F=\frac{1}{n(n+1)}\left[Tr\left(U^{\dagger}U \right) +\vert Tr\left(U_0^{\dagger}U\right)\vert^2 \right],
\end{equation}
where $n$ is the Hilbert space dimension, $U$ is the generated gate, and $U_0$ is the desired gate, we compute that the {\sc cnot} gate generated in Eq.~\ref{eq:cnot} with the parameters \{$A_{(a)},\tau_{(a)}$\} has a $99.994\%$ fidelity and a gate time $\tau_{(a)}\approx 45\mathrm{ns}$ (this does not take into account the gate time for the single-qubit gates $K_i^{(a)}$). On the other hand, the second set of parameters \{$A_{(b)},\tau_{(b)}$\} generates a {\sc cnot} gate with $99.997\%$ fidelity and a gate time $\tau_{(b)}\approx 124\mathrm{ns}$. Even though this second {\sc cnot} gate has a longer gate time, the control pulse amplitude is relatively small, corresponding to a maximum Rabi frequency approximately equal to $9\mathrm{MHz}$ (at resonance, the Rabi frequency for the left qubit is equal to $B_{y,L}^{1}(t)/2$), and thus easier to implement in the lab---the up-to-date Rabi frequencies for single-spin qubits in Si quantum dots reported in the literature range from 0.5MHz~\cite{Veldhorst2015a} up to 35MHz\cite{Takeda2016,Yoneda2017}. This shows the versatility of our scheme which, by tweaking parameters, produces two-qubit gates that conform with the experimental constraints; for example, for systems with short coherence times, shorter gates can be obtained to increase circuit depth, for long coherence times, where gate duration is less of a concern, our approach also offers gates with high fidelities.

\begin{figure}[!tbp]
  \centering
  \includegraphics[trim=0cm 3cm 0cm 3cm, clip=true,width=8.5cm, angle=0]{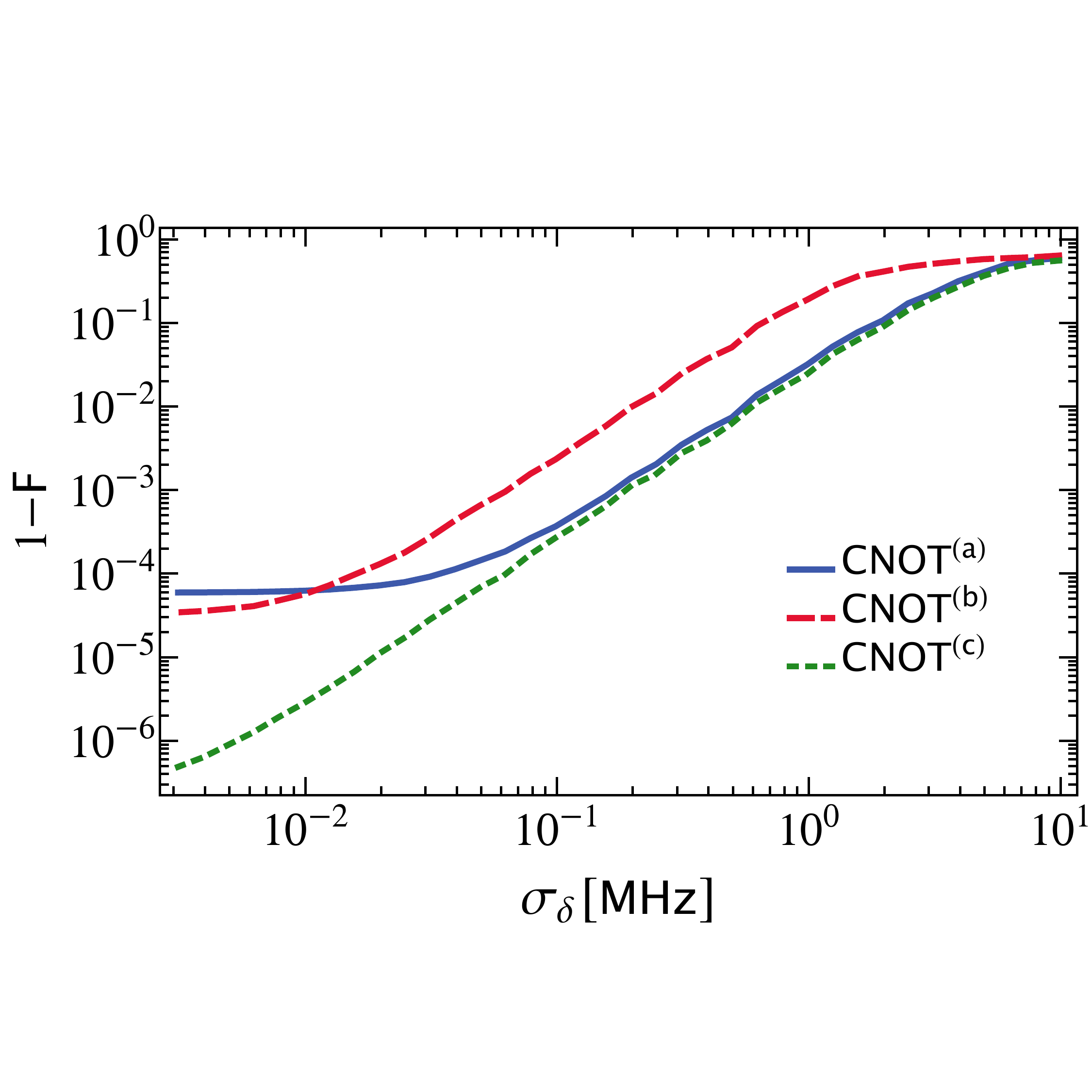}
  \caption{Infidelity of the three {\sc cnot}$^{(j)}$ gates, where $j=a,b,c$ corresponds to the control pulses depicted in Fig.~\ref{fig:2}. The blue (solid) curve corresponds to a gate of duration 45ns, the red (long-dashed) curve corresponds to a gate of duration 124ns, and the green (short-dashed) curve corresponds to the composite gate, with duration 92ns (it does not take into account the single-qubit gate times).}\label{fig:3}
\end{figure}

Now we simulate the gate-fidelity loss caused by the low-frequency component of $1/f$ charge noise, which is approximated as quasistatic during the operation time. Comparisons between theory and experiment have shown that the quasistatic approximation works well, provided the pulses are not too long~\cite{Martins2016,Reed2016}. This quasistatic noise drives fluctuations in the exchange coupling, represented by $\delta J$, as well as in the Zeeman shift terms: $\delta E_z^1$ and $\delta \Delta E_z^1$, respectively. Consequently, we form the noisy Hamiltonian by replacing the exchange coupling and Zeeman shift terms ($J$, $E_z^1$, $\Delta E_z^1$) with their respective perturbed forms: $J+\delta J$, $E_z^1+\delta E_z^1$, and $\Delta E_z^1+\delta\Delta E_z^1$. Then the infidelity is averaged over the quasistatic noise by independently sampling the three random variables, $\delta J$, $\delta E_z^1$, and $\delta\Delta E_z^1$, over a normal distribution of standard deviation $\sigma_\delta$ and zero mean, with the average being taken over 500 samples for each value of $\sigma_\delta$. The infidelities for both {\sc cnot} gates generated with Eq.~\eqref{eq:cnot} are shown in Fig.~\ref{fig:3}. Moreover, we can generate a gate locally equivalent to $\sqrt{\textsc{cnot}}$ that can be used as part of a simple two-piece pulse~\cite{Hill2007,Calderon-Vargas2016} that can correct some of the leading error and generate a {\sc cnot} with higher fidelity. Using the same ansatz for $\chi$ given in Eq.~\eqref{eq:ansatz} with the parameters $A_{(c)}=75.95269$ and $\tau_{(c)}=5.67638/J$, we generate a gate equivalent to $\sqrt{\textsc{cnot}}$ up to single-qubit unitaries with a gate time of 46 ns. The control pulse that generates such a gate is shown in Fig.~\hyperref[fig:2]{\ref*{fig:2}(c)}. Furthermore, the maximum amplitude of the control pulse is also relatively small in this case and within the range accessible in recent experiments~\cite{Zajac2018}. The two-piece pulse is~\cite{Calderon-Vargas2016}:
\begin{equation}\label{eq:two-piece-cnot}
\begin{aligned}
\left(\begin{smallmatrix}
  0 & 1 & 0 & 0 \\
  1 & 0 & 0 & 0 \\
  0 & 0 & 1 & 0 \\
  0 & 0 & 0 & 1
\end{smallmatrix}\right)\approx& K_1^{(c)}U_{int}(\tau_{(c)})\kappa_2^{(c)}\\
&\sigma_X^R\otimes\sigma_Y^L \\
&\kappa_1^{(c)}U_{int}(\tau_{(c)}) K_2^{(c)},
\end{aligned}
\end{equation}
where $K_i^{(c)}$, $\kappa_i^{(c)}$ are Kronecker products of single-qubit gates (see Appendix \ref{app:local gates k1, k2}), and $\sigma_X^R\otimes\sigma_Y^L$ are $\pi$ rotations around the $x$- and $y$-axes for the right (R) and left (L) qubits, respectively~\cite{Note}. The total duration of the two-piece pulse {\sc cnot} will depend on the gate times of each of the single-qubit gates that accompany the two two-qubit gates. Figure~\ref{fig:3} shows the infidelity of each of the three {\sc cnot} gates as functions of the charge noise strength. For noise with standard deviation less than 100 kHz, the two-piece pulse {\sc cnot} outperforms the other gates, but for higher noise strength its performance matches the control pulse shown in Fig.~\hyperref[fig:2]{\ref*{fig:2}(a)}. In the absence of noise, the {\sc cnot} in Eq.~\eqref{eq:two-piece-cnot} has a 99.9999\% fidelity. Moreover, for a typical 1\% noise level ($\approx 200\mathrm{KHz}$) all three {\sc cnot} gates have fidelities of at least $99\%$.

\section{{\sc cnot} gate generated by a square pulse}\label{sec:square pulse}
In the previous section we have shown that perfect entangling gates can be generated with smooth pulses in a relatively short time. Now we change our focus to simpler square pulses, which, due to their larger area per gate time, should generate entangling gates in much shorter times. In particular, we are interested in determining what type of entangling gates can be generated with a single square pulse. To this end, we consider the nonlocal, i.e. entangling, properties of a general two-qubit gate $U$, which are uniquely characterized by a set of three real numbers termed ``local invariants''~\cite{Makhlin2002,Zhang2003}. These invariants are related to the coefficients of the characteristic polynomial of the unitary and symmetric matrix $M(U)$, defined as $M(U)=(Q^{\dagger}UQ)^TQ^{\dagger}UQ$ where $Q$ denotes the transformation matrix from the logical basis to the magic basis~\cite{Hill1997}. In the magic basis any single-qubit operation is represented by a real orthogonal matrix, and thus the spectrum of $M$ is invariant under local operations. The set of local invariants are:
\begin{equation}\label{eq:invariants}
\begin{aligned}
G_1 & =\mathrm{Re}\left[\frac{\mathrm{tr}^2[M(U)]}{16 \det U}\right],\\
G_2 & =\mathrm{Im}\left[\frac{\mathrm{tr}^2[M(U)]}{16 \det U}\right],\\
G_3 & =\frac{\mathrm{tr}^2[M(U)]-\mathrm{tr}[M^2(U)]}{4\det U}.
\end{aligned}
\end{equation}
If two two-qubit gates are equivalent up to single-qubit operations, then they have the same set of local invariants, e.g. {\sc cnot} and {\sc cz} are equivalent up to local operations and, indeed, they share the same set of local invariants: $\{G_1=0,G_2=0,G_3=1 \}$.

\begin{figure}[!tbp]
  \centering
  \includegraphics[trim=0cm 2cm 0cm 2cm, clip=true,width=7.8cm, angle=0]{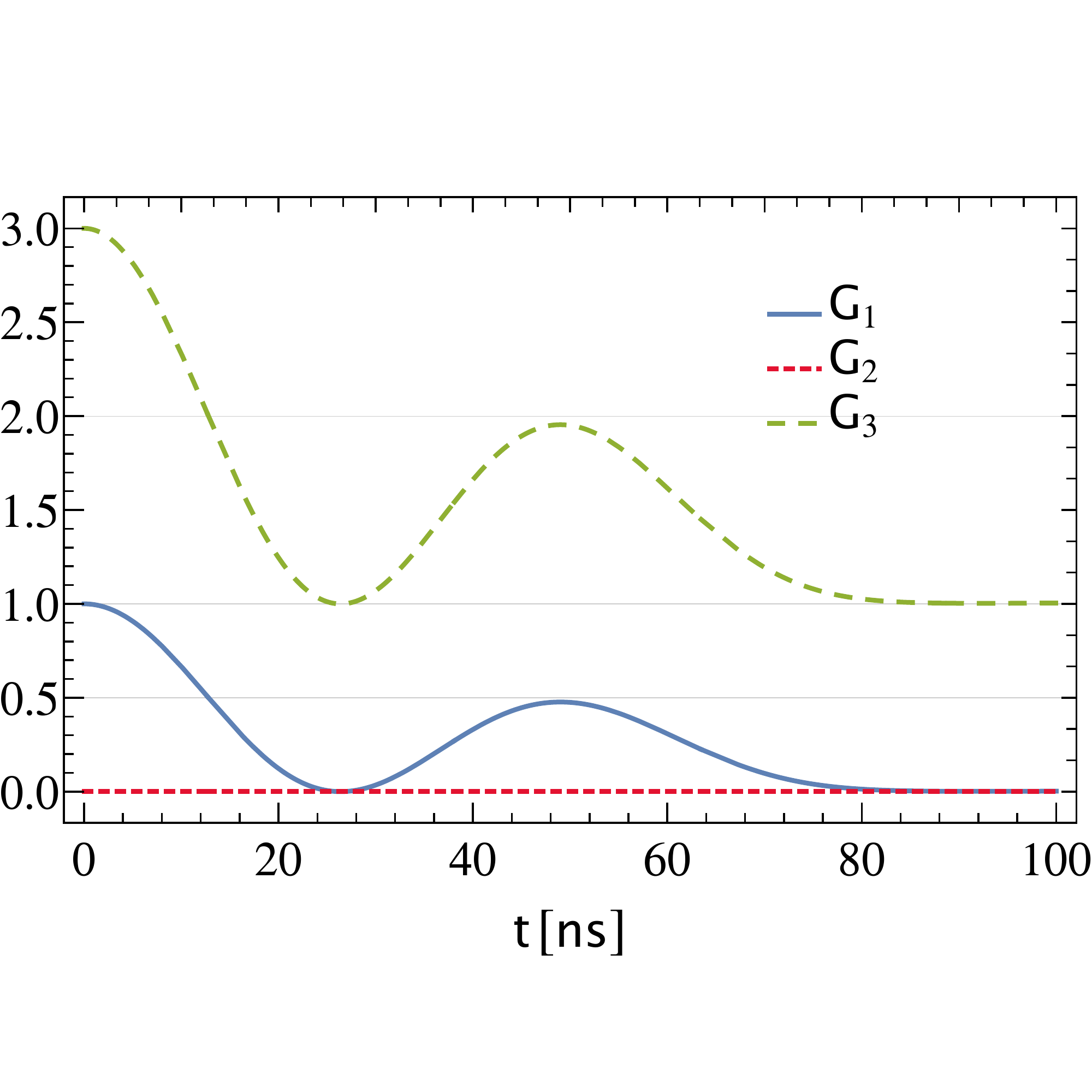}
  \caption{Local invariants $\{G_1,G_2,G_3\}$ of the system's evolution operator $U_{int}(t)$ (with square control pulse) as functions of time. At $\tau_{(1)}=26.445\mathrm{ns}$ ($\{G_1=0,G_2=0,G_3=1\}$) the evolution operator is locally equivalent to a {\sc cnot} gate, and at $\tau_{(2)}=12.8\mathrm{ns}$ ($\{G_1=0.5,G_2=0,G_3=2\}$) the evolution operator is locally equivalent to a $\sqrt{\textsc{cnot}}$ gate.}\label{fig:4}
\end{figure}

Using the Hamiltonian in the rotating frame, Eq.~\eqref{eq:Hamiltonian_rot}, with constant driving amplitude $B_{y,L}^1=B_{y,R}^1=J/2$ and driving at resonance frequency $\omega=\omega_1=E_z+E_z^1-(\Delta E_z+\Delta E_z^1-J+J^2/2(\Delta E_z+\Delta E_z^1))/2$, we calculate the local invariants of the evolution operator. Figure~\ref{fig:4} shows the set $\{G_1,G_2,G_3 \}$ versus time, where it is evident that at $t\approx26\mathrm{ns}$ the evolution operator is locally equivalent to a {\sc cnot} gate. Therefore, using the interaction picture Hamiltonian~\eqref{eq:Hamiltonian_int} with a square pulse of amplitude $B_{y,L}^1/(2\pi)=9.85\mathrm{MHz}$ and gate time $\tau_{(1)}=26.445\mathrm{ns}$ we get a {\sc cnot} gate:
\begin{equation}\label{eq:cnot_square}
\left(\begin{smallmatrix}
  0 & 1 & 0 & 0 \\
  1 & 0 & 0 & 0 \\
  0 & 0 & 1 & 0 \\
  0 & 0 & 0 & 1
\end{smallmatrix}\right)\approx K_1^{(1)}U_{int}(\tau_{(1)}) K_2^{(1)},
\end{equation}
where, again, $K_i^{(1)}$ are Kronecker products of single-qubit gates (see Appendix \ref{app:local gates k1, k2}). The {\sc cnot} gate generated in Eq.~\eqref{eq:cnot_square} has a fidelity of 99.999\%. Although we find that considerably faster gates are possible when square pulses are used, the experimental challenges with creating such ideal waveforms may lead to additional pulse generation errors, and the smooth pulses of the previous section may ultimately yield better results in practice.

In Sec.~\ref{sec:cnot gate}, we use a gate locally equivalent to $\sqrt{\textsc{cnot}}$ as part of a two-piece pulse sequence, Eq.~\eqref{eq:two-piece-cnot}, that corrects some of the leading error and generates a high-fidelity {\sc cnot} gate. Similarly, using the local invariants of the $\sqrt{\textsc{cnot}}$ gate (i.e. $\{G_1=0.5,G_2=0,G_3=2\}$), we see in Fig.~\ref{fig:4} that the system's evolution operator also generates a gate locally equivalent to $\sqrt{\textsc{cnot}}$ at $\tau_{(2)}=12.8\mathrm{ns}$. Accordingly, the two-piece pulse sequence is:
\begin{equation}\label{eq:two-piece-cnot-2}
\begin{aligned}
\left(\begin{smallmatrix}
  0 & 1 & 0 & 0 \\
  1 & 0 & 0 & 0 \\
  0 & 0 & 1 & 0 \\
  0 & 0 & 0 & 1
\end{smallmatrix}\right)\approx& K_1^{(2)}U_{int}(\tau_{(2)})\kappa_2^{(2)}\\
&\sigma_X^R\otimes\sigma_X^L \\
&\kappa_1^{(2)}U_{int}(\tau_{(2)}) K_2^{(2)},
\end{aligned}
\end{equation}
where $K_i^{(2)}$, $\kappa_i^{(2)}$ are Kronecker products of single-qubit gates (see Appendix \ref{app:local gates k1, k2}), and $\sigma_X^R\otimes\sigma_X^L$ are $\pi$ rotations around the $x$-axis for both qubits. This pulse sequence generates a {\sc cnot} gate with $99.9999\%$ fidelity in the absence of noise.

\begin{figure}[!tbp]
  \centering
  \includegraphics[trim=0cm 3cm 0cm 3cm, clip=true,width=8.5cm, angle=0]{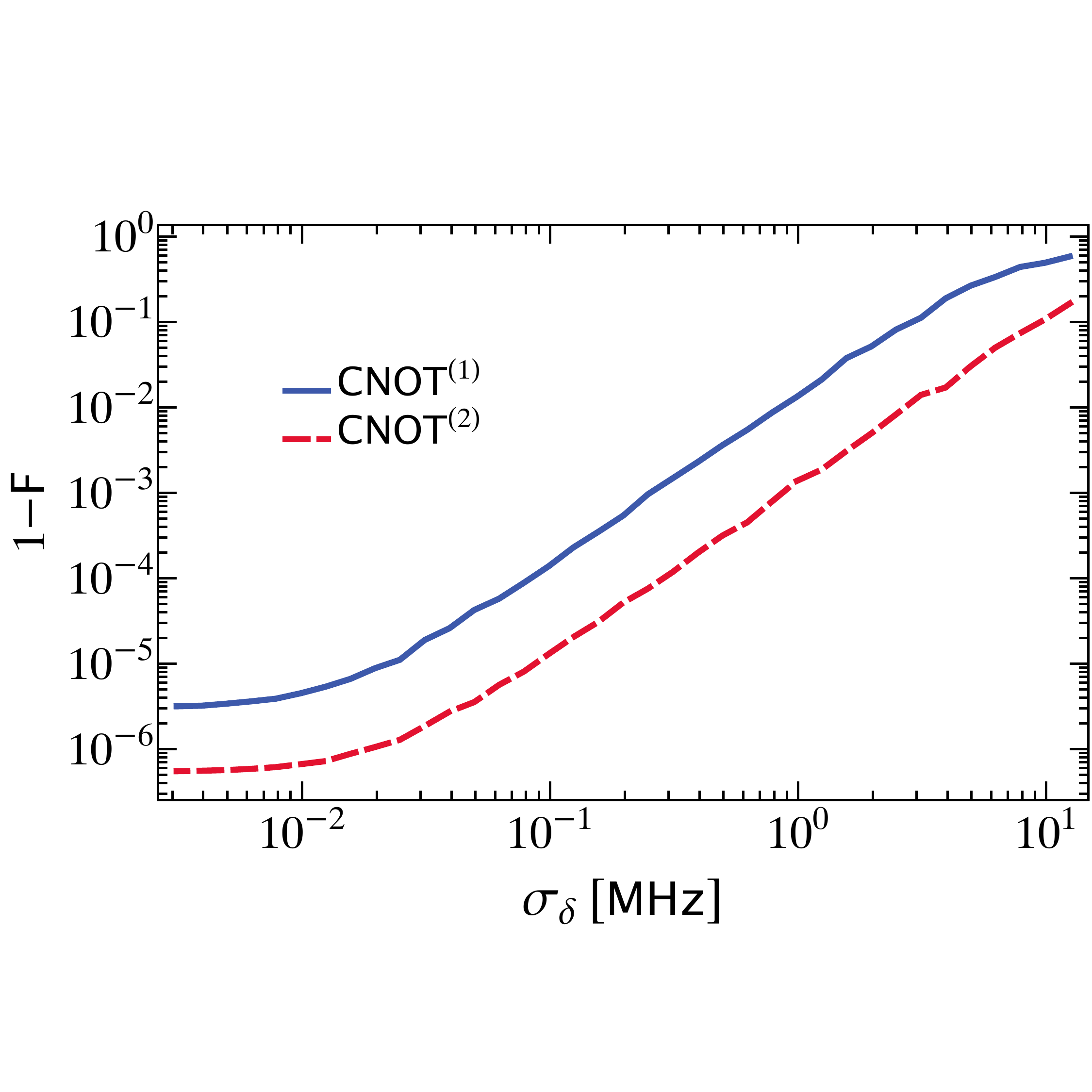}
  \caption{Infidelity of the {\sc cnot} gates vs noise strength. The blue (solid) curve corresponds to a single-shot {\sc cnot} gate, Eq.~\eqref{eq:cnot_square}, while the red (dashed) curve corresponds to a composite pulse gate, Eq.~\eqref{eq:two-piece-cnot-2}.}\label{fig:5}
\end{figure}

Next, we calculate the average gate infidelity of the {\sc cnot} gates, Eqs.~\eqref{eq:cnot_square} and \eqref{eq:two-piece-cnot-2}, versus noise strength, as shown in Fig.~\ref{fig:5}. The noise is assumed to be quasistatic and causing fluctuations in the exchange coupling $\delta J$ and both Zeeman shift terms $\delta E_z^1$ and $\delta \Delta E_z^1$. The infidelity is calculated by independently sampling the three random variables over a normal distribution of standard deviation $\sigma_\delta$ and zero mean. Its average is taken over 500 samples for each value of $\sigma_\delta$. Figure \ref{fig:5} shows that the two-piece pulse increases the {\sc cnot} fidelity by at least an order of magnitude. Furthermore, for a typical 1\% noise level ($\approx 200\mathrm{KHz}$) both {\sc cnot} gates have fidelities larger than $99.9\%$.

\section{{\sc $\theta$-cphase} and {\sc cz} gates based on sech pulses}\label{sec:cphase and cz gates}
{\sc cz} and {\sc cphase} gates naturally appear in certain quantum algorithms such as the quantum Fourier transform and in quantum simulation. To achieve shallower circuit depth, it is therefore of interest to be able to directly implement these gates instead of decomposing them in terms of {\sc cnot}s. In previous works~\cite{Economou2015,Economou2012,Economou2006}, it has been demonstrated that by using a hyperbolic secant (sech) control pulse, $\Omega(t)=\Omega_0 \mathrm{sech}(\sigma t)$ with $\Omega_0/\sigma=\mathrm{integer}$, a cyclic evolution can be induced, and thus local phase gates are easily implemented. However, the method used in those works involves an auxiliary excited state and, therefore, it is not directly applicable to the current case. Here we modify the aforementioned method such that the evolution is within the logical space. 

To illustrate the basics of the method, we apply a sech pulse in a general two-level system, whose Hamiltonian in the rotating frame is given by
\begin{equation}
\mathcal{H}(t)=\left(
\begin{array}{cc}
 -\Delta/2 & \Omega(t)  \\
 \Omega(t) & \Delta/2  \\
\end{array}
\right).
\end{equation}
The control pulse $\Omega(t)=\sigma\mathrm{sech}(\sigma t- n\pi/2)$, with $n$ being a positive real number, generates a general phase gate $\mathcal{U}=diag\{e^{-i\theta},e^{i\theta}\}$ with gate time $\tau=n\pi/\sigma$. The induced phase is given by
\begin{equation}
\theta=-2\arctan\left[\frac{\sigma}{\Delta}\right]- \frac{\Delta\tau}{2}.
\end{equation}
Here the precision of the induced phase $\theta$ is proportional to $n$.

We can easily extend this method to a two-qubit system with a Hamiltonian
\begin{equation}\label{eq:general two-qubit hamiltonian}
\mathcal{H}(t)=\left(
\begin{array}{cccc}
 -\Delta_1/2 & \Omega(t) & 0 & 0  \\
 \Omega(t) & \Delta_1/2 & 0 & 0  \\
 0 & 0 & -\Delta_2/2 & \Omega(t) \\
 0 & 0 & \Omega(t) & \Delta_2/2
\end{array}
\right),
\end{equation}
which is equivalent to the one in Eq.~\ref{eq:Hamiltonian_rot} if we again drop the terms involving $B_{y,R}^1\delta_\pm^{(2)}$ and use $\delta_\pm^{(1)}\approx 1$. In order to generate a generalized {\sc $\theta$-cphase} gate, defined as $\widetilde{\text{\sc $\theta$-cphase}}=diagonal\{e^{i\theta_1},e^{i\theta_2},e^{i\theta_3},e^{i\theta_4}\}$ with $\theta_1-\theta_2-\theta_3+\theta_4=\pm \theta $, where $\theta$ is the induced phase on the target qubit, we use the same sech pulse $\Omega(t)=\sigma\mathrm{sech}(\sigma t- n\pi/2)$. The resulting phase $\theta$ is now given by
\begin{equation}\label{eq:general theta equation}
\frac{\theta}{2}=-2\arctan\left[\frac{\sigma}{\Delta_1}\right]+2\arctan\left[\frac{\sigma}{\Delta_2}\right]-\frac{(\Delta_1-\Delta_2)\tau}{2},
\end{equation}
with a gate time $\tau=n\pi/\sigma$.

\begin{figure}[t]
  \centering
  \includegraphics[trim=1.7cm 4cm 0cm 4cm, clip=true,width=9.5cm, angle=0]{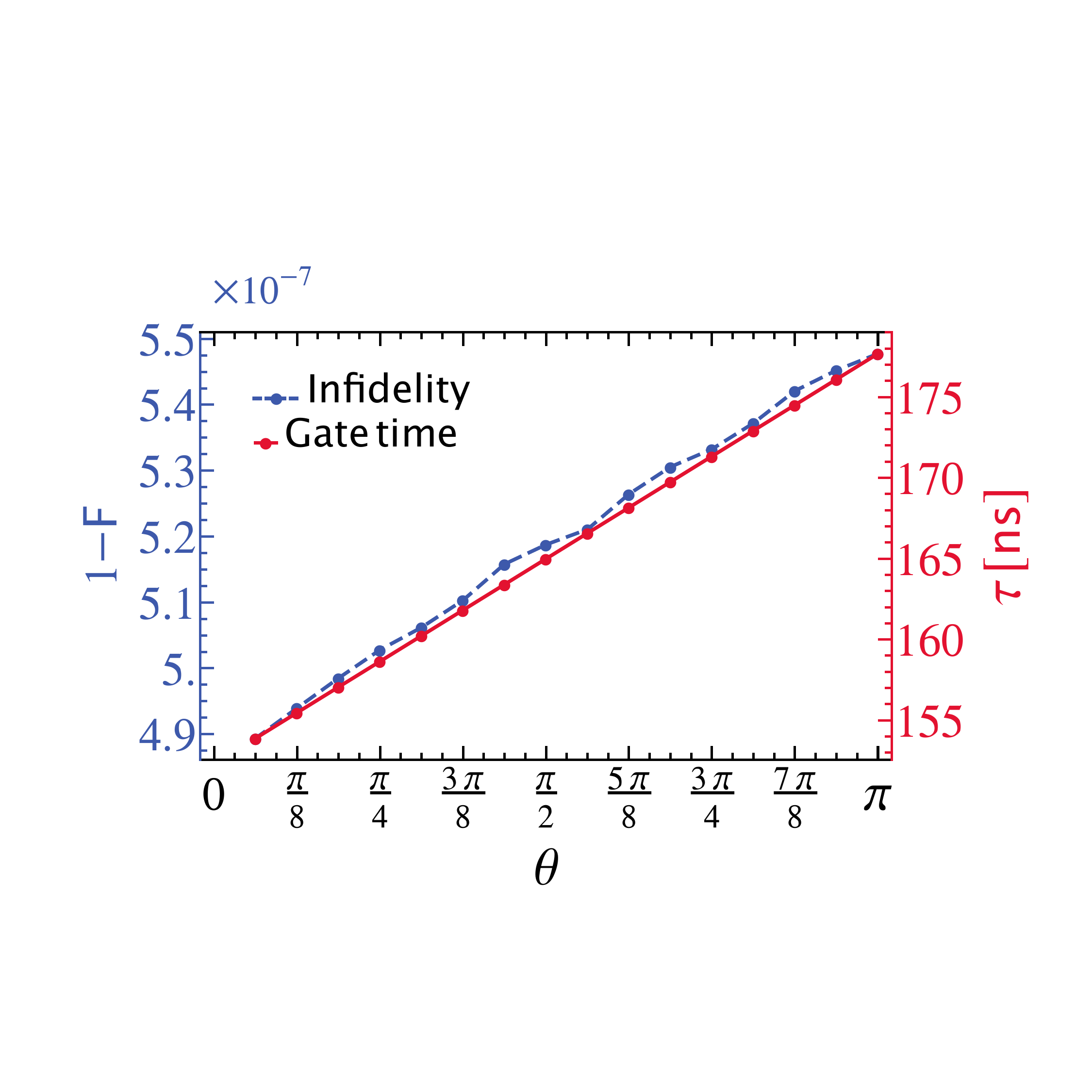}
  \caption{Infidelities and gate times for several {\sc $\theta$-cphase} gates. The abscissa gives the magnitude of the phase angle of the {\sc cphase} gate. The peak amplitudes of the control pulses, see Fig.~\hyperref[fig:7]{\ref*{fig:7}(a)}, of these gates vary between 39MHz (for $\theta=\pi/16$) and 34MHz (for $\theta=\pi$).}\label{fig:6}
\end{figure}

In order to generate a generalized {\sc $\theta$-cphase} gate with the spin Hamiltonian, we compare Eqs.~\ref{eq:Hamiltonian_rot} and \ref{eq:general two-qubit hamiltonian}, and define $\alpha=\Delta_1/J$, where $\Delta_1=\omega-\omega_1$ and $\omega_1$ is the resonance frequency of the diagonal block $S_1$. Moreover, we define $m$ to be $\sigma/J$ and, since the difference between the $S_1$ and $S_2$ detunings  is equal to $-J$, we have that $\Delta_2/J=\alpha+1$. Consequently, Eq.~\ref{eq:general theta equation} becomes:
\begin{equation}\label{eq:theta=pi equation}
\frac{\theta}{2}=-2\arctan\left[\frac{m}{\alpha}\right]+2\arctan\left[\frac{m}{\alpha+1}\right]+\frac{n\pi}{2m}.
\end{equation}
Solving Eq.~\ref{eq:theta=pi equation} for $\alpha$ gives us four solutions:
\begin{align}
\alpha_\pm^{(1)}=\frac{1}{2}\left(-1 \pm\sqrt{1-4 m^2+4 m \cot \left[\frac{1}{4} \left(\frac{n \pi  }{m}-\theta \right)\right]} \right),\label{eq:alpha_equation_1}\\
\alpha_\pm^{(2)}=\frac{1}{2}\left(-1 \pm\sqrt{1-4 m^2-4 m \tan \left[\frac{1}{4} \left(\frac{n \pi  }{m}-\theta \right)\right]} \right).\label{eq:alpha_equation_2}
\end{align}
These four equations, owing to the presence of the trigonometric functions tangent and cotangent, have real solutions only when $m$ is within the following respective ranges:
\begin{align}
&x_{r}^{(1)}\leqslant m<\frac{n \pi}{4\pi r+\theta},\\
&x_{r}^{(2)}\leqslant m<\frac{n \pi}{2\pi+4\pi r+\theta},
\end{align}
where $r$ is a positive integer and $x_{r}^{(1)}$ ($x_{r}^{(2)}$) is the root of the function $1-4 x^2+4 x \cot \left[\frac{1}{4} \left(\frac{n \pi  }{x}-\theta \right)\right] $ ($1-4 m^2-4 m \tan \left[\frac{1}{4} \left(\frac{n \pi  }{m}-\theta \right)\right]$) that is smaller than and closest in magnitude to $\frac{n \pi}{4\pi r+\theta}$ ($\frac{n \pi}{2\pi+4\pi r+\theta}$). To show the efficacy of the method, Fig.~\ref{fig:6} depicts the infidelities and gate times for several {\sc $\theta$-cphase} gates that were obtained with the interaction picture Hamiltonian~\eqref{eq:Hamiltonian_int} and the parameters listed at the end of Sec.~\ref{sec:two-qubit Hamiltonian}, along with $n=3$, $\sigma=(0.9999)\frac{n \pi}{6\pi+\theta}J$, $\omega=\omega_1+\alpha_+^{(2)} J$, where $\alpha_+^{(2)}$ is given by Eq.~\ref{eq:alpha_equation_2}, and $B_{y,L}^1(t)=4\sigma\mathrm{sech}(\sigma t- n\pi/2)$. All the {\sc $\theta$-cphase} gates in Fig.~\ref{fig:6} have fidelities exceeding 99.9999\%.

\begin{figure}[t]
  \centering
  \includegraphics[trim=2.6cm 0cm 0cm 0cm, clip=true,width=10cm, angle=0]{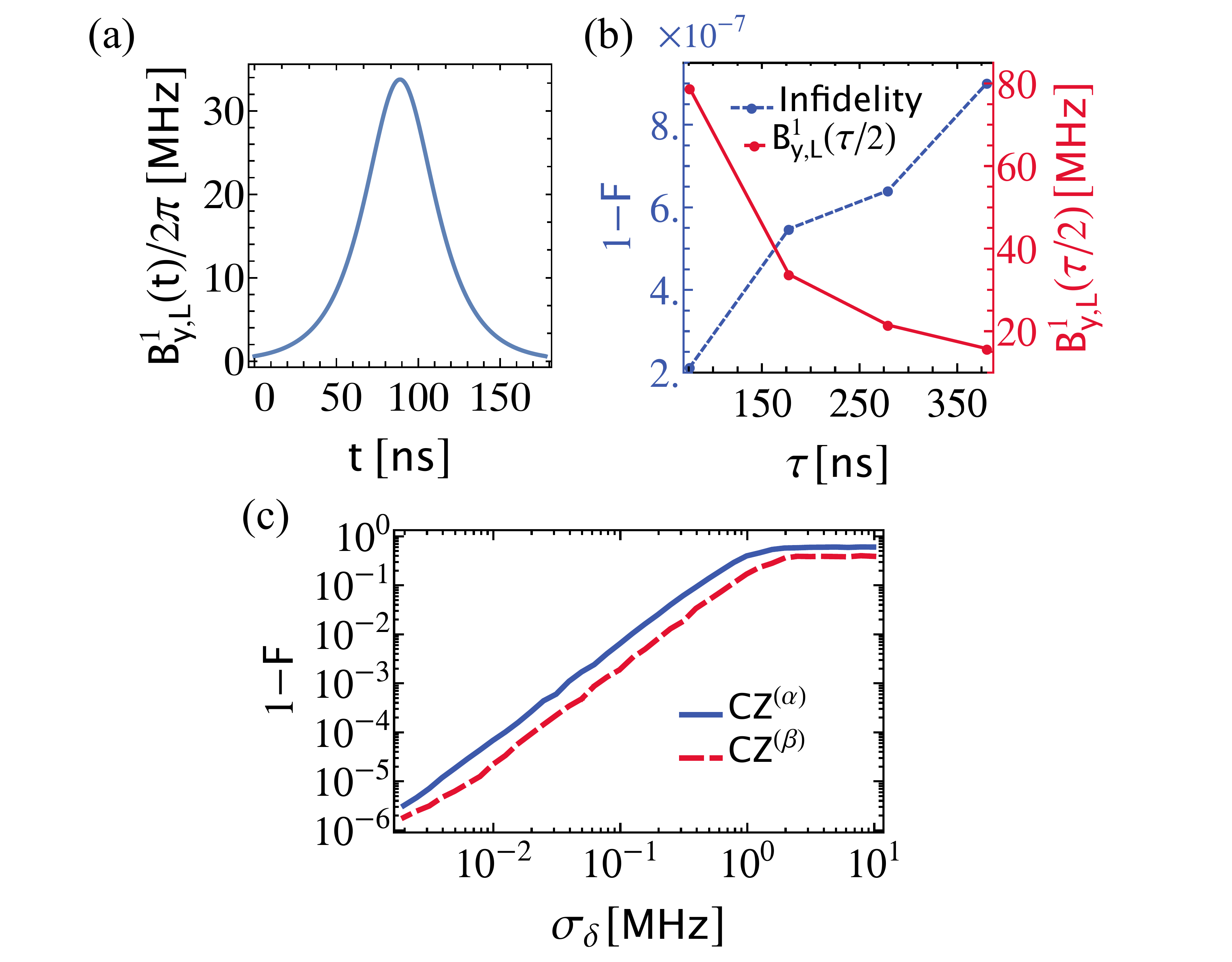}
  \caption{(a) Control pulse that generates a maximally entangling
gate locally equivalent to {\sc cz} in 178ns with 99.9999\%
fidelity. Its peak amplitude is equal to 33.8 MHz. (b) Infidelity (left ordinate, blue dashed curve) and pulse peak amplitude (right ordinate, red solid curve) versus gate time. (c) Infidelity of the {\sc cz} gates, Eqs.~\ref{eq:cz} and~\ref{eq:two-piece-cz}, versus noise strength. The blue (solid) curve corresponds to a single-shot {\sc cz} gate~\eqref{eq:cz}, while the red (dashed) curve corresponds to a composite pulse gate~\eqref{eq:two-piece-cz}.}\label{fig:7}
\end{figure}

Note that the regular $\text{\sc $\theta$-cphase}=diagonal\{1,e^{i\theta},1,1\}$ can be obtained from $\widetilde{\text{\sc $\theta$-cphase}}$ by applying subsequent single-qubit $z$-rotations and including a global phase. Moreover, after calculating the local invariants, Eq.~\ref{eq:invariants}, for $\widetilde{\text{\sc $\theta$-cphase}}$ we see that $\widetilde{\text{\sc $\theta$-cphase}}$ is maximally entangling only when $\theta_1-\theta_2-\theta_3+\theta_4=\pm \pi$, which corresponds to a generalized control-Z gate, $\widetilde{\text{\sc cz}}$~\cite{Economou2015}. Accordingly, using the same parameters employed in the calculation of the {\sc $\theta$-cphase} gates shown in Fig.~\ref{fig:6}, we generate a regular {\sc cz} gate with the interaction picture Hamiltonian:
\begin{equation}\label{eq:cz}
\left(\begin{smallmatrix}
  1 & 0 & 0 & 0 \\
  0 & -1 & 0 & 0 \\
  0 & 0 & 1 & 0 \\
  0 & 0 & 0 & 1
\end{smallmatrix}\right)\approx K_1^{(\alpha)}U_{int}(\tau_{(\alpha)}) K_2^{(\alpha)},
\end{equation}
where $K_i^{(\alpha)}$ are Kronecker products of single-qubit gates (see Appendix \ref{app:local gates k1, k2}). The {\sc cz} gate generated in Eq.~\ref{eq:cz} has a $99.9999\%$ fidelity and a gate time $\tau\approx 178\mathrm{ns}$ (this does not take into account the gate time for the single-qubit gates $K_i^{\textsc{cz}}$). Moreover, Figure~\hyperref[fig:7]{\ref*{fig:7}(a)} shows the control pulse $B_{y,L}^1(t)=4\sigma\mathrm{sech}(\sigma t- n\pi/2)$ versus time, which has a peak amplitude of 33.8 MHz. Furthermore, Fig.~\hyperref[fig:7]{\ref*{fig:7}(b)} shows that the amplitude of the sech pulse is, as expected, inversely proportional to the gate time and that the gate fidelity (99.9999\%) is slightly larger in longer gates. Therefore, a shorter gate would require a larger peak amplitude which is currently unattainable due to experimental limitations: the largest Rabi frequency reported in the literature is 35 MHz~\cite{Takeda2016,Yoneda2017} which corresponds to a pulse peak amplitude of 70 MHz.

In similar fashion to the use of $\sqrt{\textsc{cnot}}$ in Eq.~\eqref{eq:two-piece-cnot}, we use the {\sc $\pi/2$-cphase} gate as part of a two-piece pulse sequence that corrects some of the leading error and generates a {\sc cz} gate. The {\sc $\pi/2$-cphase} pulse is similar to the one depicted in Fig.~\hyperref[fig:7]{\ref*{fig:7}(a)} with a slightly larger peak amplitude $B_{y,L}^1(\tau_{(\beta)}/2)=36.4\mathrm{MHz}$ and shorter gate time $\tau_{(\beta)}=165\mathrm{ns}$. The two-piece pulse sequence is:
\begin{equation}\label{eq:two-piece-cz}
\begin{aligned}
\left(\begin{smallmatrix}
  1 & 0 & 0 & 0 \\
  0 & -1 & 0 & 0 \\
  0 & 0 & 1 & 0 \\
  0 & 0 & 0 & 1
\end{smallmatrix}\right)\approx& K_1^{(\beta)} U_{int}(\tau_{(\beta)})\kappa_2^{(\beta)}\\
&\sigma_X^R\otimes\sigma_Y^L \\
&\kappa_1^{(\beta)}U_{int}(\tau_{(\beta)}) K_2^{(\beta)},
\end{aligned}
\end{equation}
where $K_i^{(\beta)}$ and $\kappa_i^{(\beta)}$ are Kronecker products of single-qubit gates (see Appendix \ref{app:local gates k1, k2}). This pulse sequence generates a {\sc cz} gate with $99.9999\%$ fidelity in the absence of noise.

Finally, we calculate the average gate infidelity of the previous {\sc cz} gates, this is depicted in Fig.~\hyperref[fig:7]{\ref*{fig:7}(c)}. The noise is assumed to be quasistatic and affecting the exchange coupling $\delta J$ and both Zeeman shift terms $\delta E_z^1$ and $\delta \Delta E_z^1$. We independently sample the three random variables ($\{\delta J,\delta E_z^1,\delta \Delta E_z^1\} $) over a normal distribution of standard deviation $\sigma_\delta$ and zero mean. Its average is taken over 500 samples for each value of $\sigma_\delta$. As seen in Fig.~\hyperref[fig:7]{\ref*{fig:7}(c)}, the two-piece {\sc cz} gate has slightly better fidelity than the single-shot {\sc cz} gate, for a typical 1\% noise level ($\approx 200\mathrm{KHz}$) the two-piece pulse produces a {\sc cz} gate with an average fidelity larger than $99\%$.

\section{Conclusions}\label{sec:Conclusion}
Using a protocol to analytically design the unitary evolution of an arbitrary two-level system, we have presented  a set of control pulses that generate high-fidelity {\sc cnot}, {\sc cphase}, and {\sc cz} gates in the system formed by two electrons confined in a silicon double quantum dot. With these smooth control pulses, and using real parameters from the experiment reported in Ref.~\onlinecite{Zajac2018}, we have predicted, in the absence of noise, single-shot {\sc cnot} gates with more than $99.99\%$ fidelity in times as short as $45\mathrm{ns}$, and a {\sc cz} gate with $99.9999\%$ fidelity and gate time of $178\mathrm{ns}$. Moreover, after calculating the local invariants of the evolution operator we have shown that a simple square pulse can generate a {\sc cnot} gate in less than 27ns and with a fidelity of $99.999\%$. In addition, we have shown that high-fidelity ($\sim 99.9999\%$) {\sc cphase} gates with arbitrary phase angles can be implemented with simple sech pulses. Furthermore, using a simple two-piece pulse sequence interrupted by single-qubit gates, we have appreciably improved the fidelity of {\sc cnot} and {\sc cz} gates in the presence of quasistatic noise. In conclusion, we have presented a set of control pulses for fast high-fidelity two-qubit gates, whose predicted performances do not have parallel in other theoretical and experimental works, and thus they are immediately relevant for current experiments on exchange-coupled single-spin qubits in silicon quantum dots.

\section{Acknowledgments}
We are grateful to J.~R.~Petta for useful discussions and feedback on the manuscript. This work is supported by the Army Research Office (W911NF-17-0287) and the U.S. Office of Naval Research (Grant No. N00014-17-1-2971).
\begin{appendix}

\section{Single-qubit gates for {\sc cnot} and {\sc cz} gates}\label{app:local gates k1, k2}
Single-qubit gates for the {\sc cnot} gates, Eqs.~\eqref{eq:cnot} and~\eqref{eq:two-piece-cnot}, in Sec.~\ref{sec:cnot gate} are:
\begin{equation}
\begin{aligned}
K_1^{(a)}&=\exp\left[i(0.00564\sigma_X-0.006156\sigma_Y+0.507546\sigma_Z)\right]_R\\
&\otimes\exp\left[i(-0.107076\sigma_X+0.091219\sigma_Y-0.663155\sigma_Z )\right]_L,\\
K_2^{(a)}&=\exp\left[i(0.005831\sigma_X+0.006493\sigma_Y-0.320524\sigma_Z)\right]_R\\
&\otimes\exp\left[i(0.104122\sigma_X-0.008626\sigma_Y+0.034753\sigma_Z )\right]_L,\\
K_1^{(b)}&=\exp\left[i(0.000429\sigma_X-0.001655\sigma_Y-0.836667\sigma_Z)\right]_R\\
&\otimes\exp\left[i(-0.101941\sigma_X-0.204093\sigma_Y+1.37345\sigma_Z )\right]_L,\\
K_2^{(b)}&=\exp\left[i(0.00087\sigma_X+0.001517\sigma_Y+1.018818\sigma_Z)\right]_R\\
&\otimes\exp\left[i(0.153317\sigma_X+0.090985\sigma_Y-0.816242\sigma_Z )\right]_L,\\
K_1^{(c)}&=\exp\left[i(0.003253\sigma_X-0.000834\sigma_Y-1.120708\sigma_Z)\right]_R\\
&\otimes\exp\left[i(0.356721\sigma_X-0.003073\sigma_Y-0.400727\sigma_Z )\right]_L,\\
K_2^{(c)}&=\exp\left[i(0.04293\sigma_X+1.57321\sigma_Y-0.000016\sigma_Z)\right]_R\\
&\otimes\exp\left[i(-0.174612\sigma_X+0.106183\sigma_Y-1.42782\sigma_Z )\right]_L,\\
\kappa_1^{(c)}&=\exp\left[i(0.746971\sigma_X+1.379465\sigma_Y+0.002712\sigma_Z)\right]_R\\
&\otimes\exp\left[i(-0.248607\sigma_X-0.622127\sigma_Y-0.313053\sigma_Z )\right]_L,\\
\kappa_2^{(c)}&=\exp\left[i(0.208987\sigma_X-1.553846\sigma_Y+0.000769\sigma_Z)\right]_R\\
&\otimes\exp\left[i(0.328546\sigma_X+0.577004\sigma_Y-0.23246\sigma_Z )\right]_L,
\end{aligned}
\end{equation}
where the subindices $L$ and $R$ refer to the left and right qubits, respectively.

The single-qubit gates for the {\sc cnot} gate, Eqs.~\eqref{eq:cnot_square} and~\eqref{eq:two-piece-cnot-2}, in Sec.~\ref{sec:square pulse} are:
\begin{equation}
\begin{aligned}
K_1^{(1)}&=\exp\left[i(0.031741\sigma_X-0.024265\sigma_Y+2.143595\sigma_Z)\right]_R\\
&\otimes\exp\left[i(-0.157713\sigma_X-0.835226\sigma_Y+0.22501\sigma_Z )\right]_L,\\
K_2^{(1)}&=\exp\left[i(-0.010813\sigma_X-0.011452\sigma_Y-0.818712\sigma_Z)\right]_R\\
&\otimes\exp\left[i(0.402313\sigma_X+0.941847\sigma_Y+0.138859\sigma_Z )\right]_L,\\
K_1^{(2)}&=\exp\left[i(1.23768\sigma_X+0.97176\sigma_Y-0.02182\sigma_Z)\right]_R\\
&\otimes\exp\left[i(-0.0331\sigma_X-0.88781\sigma_Y-0.06691\sigma_Z )\right]_L,\\
K_2^{(2)}&=\exp\left[i(-0.00418\sigma_X-0.01482\sigma_Y-0.3255\sigma_Z)\right]_R\\
&\otimes\exp\left[i(0.64823\sigma_X-0.68884\sigma_Y-0.49805\sigma_Z )\right]_L,\\
\kappa_1^{(2)}&=\exp\left[i(-0.003159\sigma_X+0.013948\sigma_Y+0.233523\sigma_Z)\right]_R\\
&\otimes\exp\left[i(-0.007219\sigma_X-0.106195\sigma_Y+0.204111\sigma_Z )\right]_L,\\
\kappa_2^{(2)}&=\exp\left[i(0.001186\sigma_X-0.015079\sigma_Y+0.027787\sigma_Z)\right]_R\\
&\otimes\exp\left[i(0.060973\sigma_X-0.124889\sigma_Y-1.244681\sigma_Z )\right]_L.
\end{aligned}
\end{equation}

The single-qubit gates for the {\sc cz} gate, Eqs.~\eqref{eq:cz} and~\eqref{eq:two-piece-cz}, in Sec.~\ref{sec:cphase and cz gates} are:
\begin{equation}
\begin{aligned}
K_1^{(\alpha)}&=\exp\left[i(-0.036888\sigma_X+1.571355\sigma_Y+0.00069\sigma_Z)\right]_R\\
&\otimes\exp\left[i(0.000045\sigma_X-0.000142\sigma_Y+0.34968\sigma_Z )\right]_L,\\
K_2^{(\alpha)}&=\exp\left[i(-0.626514\sigma_X+1.440206\sigma_Y-0.002372\sigma_Z)\right]_R\\
&\otimes\exp\left[i(0.000262\sigma_X+0.000279\sigma_Y-0.344756\sigma_Z )\right]_L,\\
K_1^{(\beta)}&=\exp\left[i(0.001121\sigma_X-0.00114\sigma_Y+0.529903\sigma_Z)\right]_R\\
&\otimes\exp\left[i(1.567642\sigma_X-0.09954\sigma_Y-0.000507\sigma_Z )\right]_L,\\
K_2^{(\beta)}&=\exp\left[i(-0.222578\sigma_X+1.555129\sigma_Y-0.002324\sigma_Z)\right]_R\\
&\otimes\exp\left[i(0.000321\sigma_X-0.000072\sigma_Y-0.692129\sigma_Z )\right]_L,\\
\kappa_1^{(\beta)}&=\exp\left[i(0.852858\sigma_X+1.319655\sigma_Y+0.002282\sigma_Z)\right]_R\\
&\otimes\exp\left[i(0.000107\sigma_X+0.000308\sigma_Y+0.259482\sigma_Z )\right]_L,\\
\kappa_2^{(\beta)}&=\exp\left[i(-0.697808\sigma_X+1.406964\sigma_Y-0.002295\sigma_Z)\right]_R\\
&\otimes\exp\left[i(0.000197\sigma_X+0.000242\sigma_Y+0.416163\sigma_Z )\right]_L.
\end{aligned}
\end{equation}

\end{appendix}
\bibliography{library}

\begin{thebibliography}{34}
\bibitem [{\citenamefont {{We choose to use an $\sigma_X \otimes \sigma_Y$
  operation instead, let's say}}()}]{Note}%
  \BibitemOpen
  \bibfield  {author} {\bibinfo {author} {\bibnamefont {{We choose to use an
  $\sigma_X \otimes \sigma_Y$ operation instead, let's say}}},\ }\href@noop {}
  {\bibinfo  {journal} {{an $ \sigma_X\otimes\sigma_X$ operation because the
  former corrects more leading errors}}\ }\BibitemShut {NoStop}%
\end{thebibliography}
\end{document}